\documentclass[aps,pre,twocolumn,showpacs]{revtex4}

\usepackage{graphicx}

\begin{document}

\title{Comment on ``Helical MRI in magnetized Taylor-Couette flow''}

\author{G. R\"udiger}
\address{Astrophysikalisches Institut Potsdam, An der Sternwarte 16,
D-14482 Potsdam, Germany}
\author{R. Hollerbach}
\address{Department of Applied Mathematics, University of Leeds,
Leeds, LS2 9JT, United Kingdom}

\date{\today}

\begin{abstract}
Liu {\it et al.\/} [Phys.\ Rev.\ E {\bf 74}, 056302 (2006)] have presented
a WKB analysis of the helical magnetorotational instability (HMRI), and
claim that it does not exist for Keplerian rotation profiles.  We show
that if radial boundary conditions are included, the HMRI can exist even
for rotation profiles as flat as Keplerian, provided only that at least
one of the boundaries is sufficiently conducting.

\end{abstract}


\maketitle

The helical magnetorotational instability (HMRI) \cite{R01,R02,R03,R04,
R05}, is similar to the standard magnetorotational instability (SMRI)
\cite{R06,R07,R08}, in the sense that both are mechanisms whereby
hydrodynamically stable differential rotation profiles may be destabilized
by the addition of magnetic fields.  However, the way in which they behave
in the limit of small magnetic Prandtl number Pm is very different.
Unlike the SMRI, which ceases to exist for zero Pm, the HMRI continues to
exist, with the relevant nondimensional parameters being the hydrodynamic
Reynolds number Re and the Hartmann number Ha, but the magnetic
Reynolds number $\rm Rm=Pm\cdot Re$ and the Lundquist number $\rm S=Pm^{1/2}\cdot
Ha$ (the relevant parameters for the SMRI) tending to zero along with Pm.

Given these very different scalings, one natural question to ask is whether
they exist for the same range of rotation profiles.  The SMRI is known to
operate for any outwardly decreasing profile.  In contrast, the HMRI is
more delicate; the results of \cite{R01} indicate that as one moves beyond
the Rayleigh line to ever flatter profiles, one eventually switches back to
the SMRI scalings.  Liu {\it et al.\/} \cite{R09} explored this
issue more systematically, and claim that the HMRI cannot exist for
profiles as flat as Keplerian.  Specifically, their WKB analysis indicates
that if $\Omega\sim r^n$, then the HMRI requires $n<-1.66$, thereby
excluding the Keplerian value $n=-1.5$.  [Their Eq.\ (12), with their
$Ro=n/2$.]

We do not dispute the validity of their analysis; indeed, some of
our results below are in excellent agreement with it.  However, we note
that any WKB analysis is necessarily local, and does not incorporate the
boundary conditions of the global problem.  We show here that if the
radial boundary conditions are taken into account, the HMRI can exist
even for the Keplerian value $n=-1.5$, provided only that one of the
boundaries is at least somewhat conducting.

The eigenvalue problem to be solved is
$${\rm Re}\,\gamma\,v=
      D^2v+{\rm Re}\,ik\,r^{-1}(r^2\Omega)'\,\psi+{\rm Ha}^2\,ik\,b,$$
$${\rm Re}\,\gamma\,D^2\psi=
    D^4\psi-{\rm Re}\,2ik\,\Omega\,v+{\rm Ha}^2(k^2\psi+2ik\beta r^{-2}b),$$
$$0=D^2b+ik\,v-2ik\,\beta r^{-2}\psi,$$
essentially the same as in \cite{R01}, except that we restrict
attention to the $\rm Pm\to0$ limit, ensuring that any instabilities we
obtain will necessarily be the HMRI.  (Priede {\it et al.\/} \cite{R10}
have very recently also considered this $\rm Pm\to0$ limit; some of their
results are quite relevant here, and will be discussed below.)

The boundary conditions associated with $v$ and $\psi$ are no-slip, just
as in \cite{R01}.  The boundary conditions associated with $b$ are
$$b=\epsilon(rb)'\quad{\rm at}\ \ r_{\rm i}=1,\qquad
   b=0\quad{\rm at}\ \ r_{\rm o}=2.$$
The outer boundary is therefore insulating, whereas the nature of the
inner boundary depends on $\epsilon$:  $\epsilon=0$ is insulating,
$\epsilon=\infty$ is perfectly conducting.  Intermediate values
correspond to a boundary consisting of a thin layer of relative
conductance $\epsilon$ (see for example \cite{R11,R12} for this thin
boundary layer approximation in other contexts).

$\Omega(r)$ is the rotation profile whose stability is to be
investigated.  In \cite{R01,R02} we considered Taylor-Couette profiles
of the form $\Omega=c_1+c_2/r^2$.  To facilitate comparison with Liu
{\it et al.\/}, here we will primarily consider profiles of the form
$\Omega=r^n$.  As we will see though, the two choices yield almost
identical behavior.

Figure 1 shows contour plots of the critical Reynolds number for the onset
of the HMRI, as a function of $n$ and $\beta$, and optimized over the
Hartmann number Ha and the axial wavenumber $k$.  Turning first to the
$\epsilon=0$ plot on the left, we note how increasing $\beta$ from 1 to
5 facilitates the instability, that is, allows it to exist increasingly
far beyond the Rayleigh line at $n=-2$.  Beyond $\beta\approx5$ though
the ${\rm Re}_{\rm c}$ curves become largely independent of $\beta$.  And crucially,
even the $\rm Re_c=10^6$ curve asymptotes just to the left of the Liu {\it et
al.\/} line at $n=-1.66$.  These $\epsilon=0$ results are therefore in
excellent agreement with their prediction that the HMRI only exists to
the left of this line.

\begin{figure}[t]
\includegraphics[scale=0.6]{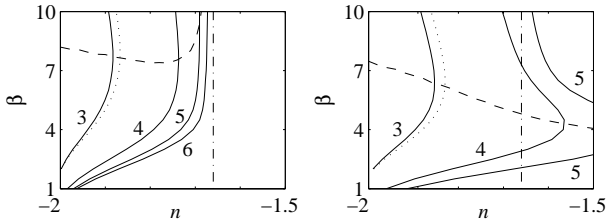}
\caption{The left panel is for $\epsilon=0$, the right for $\epsilon=
\infty$.  In each case the solid curves are contours of $\log{\rm Re_c}$ as
a function of $n$ and $\beta$, optimized over Ha and $k$.  The dashed
curves are the location where $\rm Re_c$ is optimized over $\beta$ as well.
The dotted curves show how the $10^3$ contours are shifted slightly to
the right if the $r^n$ profile is replaced by a Taylor-Couette profile
having the same $\Omega_{\rm o}/\Omega_{\rm i}$ ratio.  The dash-dotted lines denote
the $n=-1.66$ boundary, beyond which the HMRI does not exist in the Liu
{\it et al.\/} analysis.  Note how this agrees very well with our results
for $\epsilon=0$, but not at all for $\epsilon=\infty$.}
\end{figure}

\begin{figure}[t]
\includegraphics[scale=0.8]{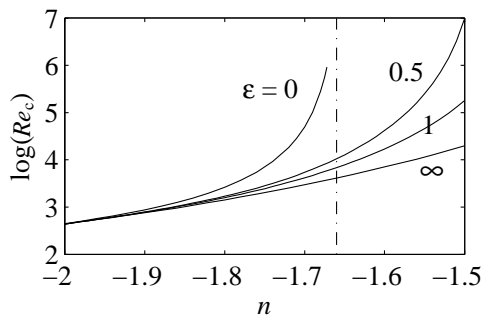}
\caption{$\log{\rm Re_c}$ as a function of $n$, optimized over $k$, Ha and
$\beta$.  Note how $\epsilon=0$ does indeed appear to reach a vertical
asymptote at the Liu {\it et al.\/} value $n=-1.66$, but $\epsilon=0.5$
is already sufficiently large to reach the Kepler value $n=-1.5$.}
\end{figure}

However, if we now turn to the $\epsilon=\infty$ plot on the right, we
note that $\rm Re_c=10^4$ already extends beyond the Liu {\it et al.\/} line,
and $Re_c=10^5$ extends beyond the Kepler line $n=-1.5$.  Simply switching
the inner boundary from insulating to conducting is sufficient to allow the
HMRI to operate even for Keplerian rotation profiles.  Similar results are
obtained if instead it is the outer boundary that is switched from
insulating to conducting.  Why the electromagnetic boundary conditions
should have such a dramatic effect is not clear, but it is certainly well
known in many other contexts, e.g. \cite{R13,R14}, that they can play a
crucial role.

Note also that exactly the same phenomenon illustrated in Fig.\ 1 is
already implicit in Fig.\ 2 of \cite{R10}, where the HMRI exists up to
$\mu\approx0.32$ for insulating boundaries, but up to $\mu\approx0.45$ for
conducting boundaries, where $\mu=\Omega_{\rm o}/\Omega_{\rm i}$.  Translating from
$\mu$ to $n$ via $\mu=2^n$, their results become $n\approx-1.64$ for
insulating boundaries, versus $n\approx-1.15$ for conducting boundaries.
That is, the Keplerian value $n=-1.5$ is accessible with conducting
boundaries, but not with insulating ones.

At first sight it would appear that their value of $n\approx-1.64$ for
insulating boundaries is already in conflict with the Liu {\it et al.\/}
limit $n<-1.66$.  In fact, this slight discrepancy is due to the
difference between the $\Omega=c_1+c_2/r^2$ profile used by \cite{R10},
and the $\Omega=r^n$ profile considered by Liu {\it et al}.  Even if $c_1$
and $c_2$ are chosen to match $\Omega_{\rm i}=1$ and $\Omega_{\rm o}=2^n$, a
Taylor-Couette profile will be somewhat steeper near $r_{\rm i}$, and
correspondingly somewhat shallower near $r_{\rm o}$.  By concentrating near the
inner boundary, the instability can then operate at somewhat larger
$\Omega_{\rm o}/\Omega_{\rm i}$ values than for an $r^n$ profile.  The dotted lines in
Fig.\ 1 quantify this effect, showing the $\rm Re_c=10^3$ curves if the $r^n$
profile is replaced by a Taylor-Couette profile, with $c_1$ and $c_2$
chosen as indicated above.  We see that qualitatively the two profiles
yield exactly the same behavior, but that the Taylor-Couette profile
extends to slightly larger values of $n$.  It was precisely to avoid this
effect, and thereby allow a direct comparison with Liu {\it et al.\/},
that we chose here to concentrate on the $r^n$ profiles.

The dashed lines in Fig.\ 1 show the locations along which $\rm Re_c$ is
optimized not only over $k$ and Ha, but over $\beta$ as well.  Fig\ 2
shows $\rm Re_c$ as a function of $n$ along these lines, now including not
just $\epsilon=0$ and $\infty$, but also $\epsilon=0.5$ and 1.  We see
that $\epsilon=0.5$ is already large enough to reach the Kepler value
$n=-1.5$.  That is, if the conductance of the inner boundary is only 1/2
that of the fluid region, the HMRI already exists for rotation profiles
as flat as Keplerian.

\end{document}